\documentclass[amsmath,amssymb,aps,eqsecnum,pra,showpacs]{revtex4}
\usepackage[dvips]{graphicx}
\usepackage{bm,bbm}
\usepackage{color}

\newtheorem{theorem}{Theorem}
\newtheorem{lemma}{Lemma}
\newtheorem{proposition}{Proposition}

\newcommand{\tr}{\,\mathrm{Tr}{ }}
\newcommand{\bra}[1]{\ensuremath{\langle#1|}}
\newcommand{\ket}[1]{\ensuremath{|#1\rangle}}

\newcommand{\one}{\mathrm{I} \! \! 1}
\newcommand{\bbR}{{\rm I\hspace{-0.8mm}R}}
\newcommand{\halmos}{\newline\vspace{3mm}\hfill $\Box$}

\begin{document}
\title{Quantum state discrimination: a geometric approach}






\author{Damian Markham}
\email{markham@phys.s.u-tokyo.ac.jp}
\affiliation{Universit\'{e} Paris 7, 175 Rue du Chevaleret, 75013 Paris, France}
\affiliation{Department of Physics, Graduate School of Science, University of
Tokyo, Tokyo 113-0033, Japan}

\author{Jaros{\l}aw Adam Miszczak}
\email{miszczak@iitis.gliwice.pl}
\author{Zbigniew Pucha{\l}a}
\email{z.puchala@iitis.gliwice.pl}
\affiliation{Institute of Theoretical and Applied Informatics, Polish Academy of
Sciences, Ba{\l}tycka 5, 44-100 Gliwice, Poland}

\author{Karol \.Zyczkowski}
\email{karol@cft.edu.pl}
\affiliation {Instytut Fizyki im. Smoluchowskiego, Uniwersytet Jagiello{\'n}ski, 
ul. Reymonta 4, 30-059 Krak{\'o}w, Poland} 
\affiliation{Centrum Fizyki Teoretycznej, Polska Akademia Nauk, Al. Lotnik{\'o}w
32/44, 02-668 Warszawa, Poland}

\date{November 27, 2007}

\begin{abstract}
We analyse the problem of finding sets of quantum states that can be
deterministically discriminated. From a geometric point of view this problem is
equivalent to that of embedding a simplex of points whose distances are maximal
with respect to the Bures distance (or trace distance). We derive upper and
lower bounds for the trace distance and for the fidelity between two quantum
states, which imply bounds for the Bures distance between the unitary orbits of
both states. We thus show that when analysing minimal and maximal distances
between states of fixed spectra it is sufficient to consider diagonal states
only. Hence considering optimal discrimination, given freedom up to unitary
orbits, it is sufficient to consider diagonal states. This is illustrated
geometrically in terms of Weyl chambers.
\end{abstract}


\maketitle



\section{Introduction}

The geometry of state space depends on the distance measure chosen.
In state discrimination, given a set of possible states, our task is to 
find out as `best' as possible which of the states we have in our possession
\cite{He76,Fu96,Ch00}. Finding an optimal procedure of unambiguous
discrimination is particularly interesting if the states analyzed
are mixed \cite{RLE03,ESH04,HB05,RL05, He07}. The usual approach to
the quantum discrimination problem is to begin by considering the
classical case and then extending to the quantum case. Different
concepts of `best' induce different measures of distinguishability
in the space of classical probability distributions. In the quantum
case, on top of the statistical uncertainty of states, even pure
states cannot be always be perfectly discriminated (if the states
are not orthogonal), meaning that one has to be careful in extending
these to the quantum setting. To do this we bring it back to the
classical setting of probability distributions by maximising over
all possible discrimination measurements. In this way the problem of
discriminating quantum states has led to several distance measures
associated with the ability to discriminate well (see e.g.\cite{FC95, FG99, 
BZ06}). In this work we would like to consider the geometry induced 
by these measures, and how the problem of state discrimination can be 
expressed geometrically.

More precisely, let ${\cal M}_N$ denote the set of mixed quantum states acting on an $N$ dimensional
Hilbert space ${\cal H}_N$. It is a convex, compact set of dimensionality $N^2-1$. Its geometric
structure depends on the metric used. The following distances are often used \cite{FG99,BZ06}
\begin{eqnarray}
D_{\rm HS}(\rho_1,\rho_2)&:=&[{\rm Tr}(\rho_1-\rho_2)^2]^{1/2}, \label{HS1} \\
D_{\rm tr}(\rho_1,\rho_2)&:=&\frac{1}{2}{\rm Tr}|\rho_1-\rho_2|, \label{trace1} \\
D_{\rm B}(\rho_1,\rho_2)&:=&\bigl( 2[1- {\rm Tr}|\sqrt{\rho_1}\sqrt{\rho_2}|] \bigr)^{1/2}, \label{Bures1}
\end{eqnarray}
denoting the Hilbert-Schmidt (HS) distance, the trace distance and
the Bures distance respectively. The latter quantity is a function
of {\sl fidelity} \cite{Jo94},
\begin{equation}
F(\rho_1,\rho_2):=[{\rm Tr}|\sqrt{\rho_1}\sqrt{\rho_2}|)]^2 \  , \label{fidel1}
\end{equation}
and the root fidelity $\sqrt{F}$, (which in some papers is also is called 
`fidelity'). The Bures and the trace distance are {\sl monotone}, and do 
not grow under the action of an arbitrary quantum operation (completely
positive, trace preserving map), while the Hilbert-Schmidt (HS)
distance is not monotone. These measures can induce different
geometries. For instance, the set ${\cal M}_2$ of mixed states of a
qubit, is equivalent to the standard Bloch--ball (the Bloch-sphere
and its interior) for the trace or HS metric, and to Uhlmann
hemisphere, $\frac{1}{2} S^3$, for the Bures distance \cite{Uh92}.
For higher $N$ the geometries induced by the HS and the trace
distance also differ.


In the following we consider systems of dimension $N$ greater or equal to two. We begin our discussion
of state discrimination by introducing the diameter of a set of quantum states. The diameter of the set
${\cal M}_N$ is independent of $N$, but it does depend on the metric used: the {\it diameter} is the
maximal distance between any two states, and it reads

\begin{equation}
D_{\rm HS}^{\rm max}=\sqrt{2}, \quad \quad
D_{tr }^{\rm max}=1, \quad \quad
D_{\rm B}^{\rm max}=\sqrt{2} ,
\label{dmax}
\end{equation}
for HS, trace and Bures distances, respectively. Any two states
separated by $D^{\rm max}$ are supported on orthogonal subspaces.
The reverse implication holds for Bures and trace distances,
\begin{equation} \mbox{supp}(\rho_1) \perp \mbox{supp}(\rho_2) \hspace{3mm}
\Leftrightarrow \hspace{3mm} D_{\rm tr}(\rho_1, \rho_2) = 1 \hspace{3mm}
\Leftrightarrow
\hspace{3mm} D_{\rm B}(\rho_1, \rho_2) = \sqrt{2} \ ,
\label{maxB}
\end{equation}
but is not true for the Hilbert-Schmidt distance for $N>2$. For
instance, the HS distance between two diagonal density matrices
$\rho_1={\rm diag}(1,0,0)$ and $\rho_2={\rm diag}(0,1/2,1/2)$ is
equal to $\sqrt{3/2}< D_{\rm HS}^{\rm max}$, although they are
supported on orthogonal subspaces. To witness an even more dramatic
example consider the Hilbert space of even dimension $N$ and two
diagonal states, $\rho_1={\rm diag}(N/2,...,N/2,0,...,0)$ and
$\rho_2={\rm diag}(0,...,0,N/2,...,N/2)$. Although they live in
orthogonal subspaces, so their Bures and trace distances are
maximal, their HS distance reads $2/\sqrt{N}$ and tends to zero in
the limit of large $N$. This indicates that when analysing problems
of distinguishability, one cannot therefore rely on the standard
Euclidean geometry induced by the Hilbert-Schmidt distance, but
rather better use Bures or trace distances.


The trace distance and the Bures distance are, in several respects,
good measures for quantifying the ability to discriminate states. In
\cite{En96} Englert introduced the notion of {\sl
distinguishability} between two quantum states and showed that it is
equal to the trace distance between them. Hence two states can be
deterministically discriminated if they can be perfectly
distinguished, so their distinguishability is equal to unity. Fuchs
and  van de Graaf found a bound between the Bures distance and the
trace distance based on the following inequality \cite{FG99}
\begin{equation}
1- \sqrt{F(\rho_1,\rho_2)} \ \le  \
D_{tr }(\rho_1,\rho_2)      \ \le \
\sqrt{1-F(\rho_1, \rho_2)}\ .
\label{bounddf}
\end{equation}
This implies that if the fidelity between both states is equal to
zero (so the states are distinguishable and their Bures distance is
maximal) their trace distance is equal to unity, and is hence
maximal. In fact, the trace distance is a simple function of the
probability to successfully discriminate two states in a single shot
measurement (optimised over all allowed quantum measurements)
\cite{FG99}. Similarly, the Bures distance can be seen as the
optimised Kullback--Leibler distance between output statistics over
all quantum measurements (again, an optimized cost function for
discrimination) \cite{FC95}.

In the special case where both density matrices are diagonal, and
read $p$ and $q$, the operators commute. Such a case is often
called classical since the  distances between quantum states
reduce then exactly to their classical analogues: The trace distance
$D_{\rm tr}(p,q)$ is the equal to the $L_1$ distance (with a
normalisation constant $1/2$) between both probability vectors; The
Bures distance reads $D_{\rm B}(p,q)=[2(1-B(p,q))]^{1/2}$, where
\begin{equation}
 B(p,q) :=   \sum_{i=1}^N \sqrt{p_i q_i}
\label{Bhat}
\end{equation}
denotes the {\sl Bhattacharyya} coefficient \cite{Bh43}, \cite{BZ06}. 
This quantity is equal to the root fidelity between any two diagonal 
states, $B(p,q)=\sqrt{F(p,q)}$, so its square $B^2$, is sometimes called 
{\sl classical fidelity} between to probability distributions.

In section  \ref{SCN: unitary orbits} we prove general bounds for the fidelity between
arbitrary two quantum states $\rho_1$ and $\rho_2$,
\begin{equation}
 B^2 ( p^{\uparrow}, q^{\downarrow})   \  \le \  F(\rho_1,\rho_2) \ \leq  \  B^2 ( p^{\uparrow}, q^{\uparrow}) \ ,
\label{fidboth}
\end{equation}
where the vectors  $p$ and $q$ represent the spectra of $\rho_1$ and $\rho_2$,
while the arrows up (down)
indicate that the eigenvalues are put in the nondecreasing (nonincreasing) order.
These results imply equivalent  bounds for the Bures distance
 \begin{equation}
%
 \sqrt { 2-2\sqrt{p^\uparrow}.\sqrt{q^\uparrow} }
 \ \le \  D_B (\rho_1, \rho_2)   \ \le  \
  \sqrt {2- 2 \sqrt{p^\uparrow}.\sqrt{q^\downarrow} } \  .
\label{burboth}
\end{equation}
Analogous bounds for the trace distance proved in the same section read 
 \begin{equation}
  D_{\rm tr} ({p^\uparrow}, {q^\uparrow} )
 \ \le \  D_{\rm tr} (\rho_1, \rho_2)   \ \le  \
  D_{\rm tr} ({p^\uparrow}, {q^\downarrow} )
 \  ,
\label{tracboth}
\end{equation}
where the symbols ${p^\uparrow}$  and ${q^\downarrow}$ denote here  diagonal density matrices  
with all eigenvalues in the increasing (decreasing)  order. 

In this paper we set out to give a geometric interpretation to the problem of state discrimination in
terms of the geometries induced by the trace and Bures distance. We begin
in section \ref{SCN: perfect disc} by giving a set of conditions on states
such that they may be perfectly discriminated. In section \ref{SCN:
geometrical consequences} we present some geometrical consequences of these
conditions and phrase the problem of state discrimination in terms of the
embedding of simplices with respect to different distance functions. In section
\ref{SCN: unitary orbits} we investigate the distance between states under
unitary orbits and its geometric interpretation, and prove the above bounds. We finish in section
\ref{SCN:conclusions} with conclusions.

\section{Perfect discrimination of states}
\label{SCN: perfect disc}

We begin by looking at some conditions on the set of states that can
be perfectly discriminated. Our condition will follow from simple
analysis of the measurements (in terms of the associated positive
operator valued measure (POVM)), and give general conditions which,
in the next section, will be used to give some geometrical
consequences of the problem.

\begin{theorem}
Two states $\rho_1$ and $\rho_2$ can be deterministically discriminated 
iff their supports do not overlap.
\end{theorem}


{\bf Proof}. Any perfect state discrimination strategy for two states $\rho_1, \rho_2$ can be written
as a three element POVM $\{A_1, A_2, A_?\}$, where the outcomes correspond to concluding it is the
state $\rho_1$, $\rho_2$ and allowing for inconclusive outcome respectively.

Note that although in general we can have far more possible outcomes
than three, this formalism does include all possible discrimination
strategies - this is because we can always group the outcomes
corresponding to state $\rho_1$ to give $A_1$, and those to state
$\rho_2$ to give $A_2$, and the remaining elements we group to give
$A_?$. The probability of success of the strategy can always be
written in terms of such POVMs, thus we can restrict ourselves to
only these three element POVMs for perfect discrimination.

The conditions on the POVM for deterministic state discrimination
are

\begin{eqnarray}
{\rm
Tr}(A_1 \rho_1) &=& 1 ~~~~~ \label{eqn: condition perfect1} \\
{\rm Tr}(A_2 \rho_2) &=& 1 ~~~~~ \label{eqn: condition perfect2}\\
 A_1 + A_2 +A_? &=& \one \label{eqn: condition completeness}\\
\one \geq A_i &\geq& 0 ~~~~~ \label{eqn: condition positivity}
\end{eqnarray}
(this is the same logic as in \cite{HMMOV06}, only without the
separability condition). The first two are necessary for perfect
state discrimination, and the last two are just the conditions for
$\{A_i\}$ to be a POVM.

Conditions (\ref{eqn: condition perfect1}) and (\ref{eqn: condition
perfect2}) imply that the elements $A_1$ and $A_2$ include
projections onto the support of $\rho_1$ and $\rho_2$ respectively.
To see this, rewrite (\ref{eqn: condition perfect1}) in the
eigenbasis of $\rho_1 = \sum_i \lambda_i |i\rangle\langle i|$ (we
extend this basis to the full space for writing $A_1$ in (\ref{eqn:
M1}))
\begin{eqnarray}
{\rm Tr}(A_1 \rho_1)&=&\sum_i\lambda_i \langle i|A_1 |i\rangle
\nonumber\\
&=& \sum_i \lambda_i q_i =1,
\end{eqnarray}
where $q_i:=\langle i|A_1 |i\rangle$ is a probability, hence $\sum_i
\lambda_i q_i\leq1$ and equality is obtained only when $q_i=1$ for
all $i$ such that $\lambda_i\neq 0$. If we also demand conditions
(\ref{eqn: condition completeness}),(\ref{eqn: condition
positivity}) the most general $A_k$ can be written
\begin{eqnarray} \label{eqn: M1}
A_k= P_k+ \sum_{i,j \notin Supp(\rho_1),Supp(\rho_2)}
\alpha_{i,j}|i\rangle\langle j|
\end{eqnarray}
where $P_k=\sum_{i\in Supp(\rho_k)} |i\rangle\langle i|$ is the
projector onto the support of state $\rho_k$. The support of a state
$\rho$, with eigen-decomposition $\rho = \sum_j \alpha_j
|j\rangle\langle j|$ is given by $P=  \sum_j |j\rangle\langle j|$ .
From here, condition (\ref{eqn: condition completeness}) clearly
says
\begin{eqnarray}
P_1+P_2 \leq \one \nonumber\\
\Rightarrow {\rm
Tr}(P_1P_2)&=&0 \nonumber\\
\Rightarrow {\rm
Tr}(\rho_1\rho_2) &=&0 \nonumber \\
\Rightarrow {\rm Tr}|\rho_1-\rho_2|/2&=&1.
\end{eqnarray}
Hence the supports have zero overlap.  \halmos

The theorem is easily extended to sets of states
$\{\rho_i\}_{i=1}^M$. 

\begin{theorem}
The states $\{\rho_i\}_{i=1}^M$ can be deterministically discriminated 
iff their supports do not overlap.
\end{theorem}


This directly leads to

\begin{proposition}
  Consider $K$ states acting on the $N$ dimensional Hilbert space, which 
  can be discriminated deterministically. Then
  \begin{equation}
	\sum_{i=1}^K {\rm rank}(\rho_i) \le N. \label{rankmax}
  \end{equation}
\end{proposition}

This proposition is clear from the theorem, but also can be derived
from the result in \cite{HMMOV06}. This is done by taking the zero
entanglement case of the main result presented there. Specifically,
the left hand inequality in equation (8) for zero entanglement,
along with equation (1) in \cite{HMMOV06} give exactly
(\ref{rankmax}).

\section{Some Geometrical Consequences}
\label{SCN: geometrical consequences}

We now look at what the above results have to say in terms of the
geometric interpretation of the problem of state discrimination. Due
to property (\ref{maxB}) the above theorem can also be formulated as
the condition that the trace (or Bures) distance between states are
maximal. This fact has an immediate geometric implication. Let us
start to work with the trace distance and denote by $\Delta_k\in
{\mathbb R}^{k}$ a  maximal regular $k-simplex$ defined by $k+1$
points with mutual trace distance between points equal to
$D_{tr}^{\rm max}=1$.

\begin{proposition}\label{prop:3}
  Let $R$ be an arbitrary convex subset of  ${\cal M}_N$.
  Assume that there exists a simplex $\Delta_k \subset R$
  and assume that $R$ does not contain $\Delta_{k+1}$.
  Then the maximal number of states of $R$
  which can be discriminated deterministically is equal to
  $k+1$.
\end{proposition}


An analogous of the Proposition \ref{prop:3} may be formulated for the
geometry induced by the Bures distance.


\begin{figure} [htbp]
  \begin{center}
	\includegraphics[width=11.0cm,angle=0]{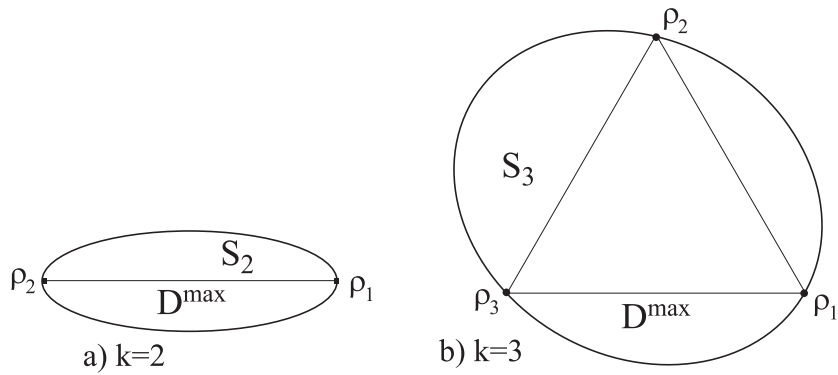}
	\caption{Set of positive operators $\rho_1,\dots \rho_k$
	with a) $k=2$ and b) $k=3$ distinguishable states which form 
	a~maximal simplex of size $k$ with side length $ D^{\rm max}$, 
	with respect to the Bures (or the trace) metric.} \label{fig1}
  \end{center}
\end{figure}

Thus the problem of finding the maximal number of distinguishable states on a certain set is equivalent
to the problem of embedding inside it a regular simplex of maximal dimensionality with the diameter
given by $D^{\rm max}$ (see Fig.~\ref{fig1}).

At this point it is worth mentioning a different quantum problem of
finding `symmetric, informationally complete positive operational
valued measures' (SIC POVM) \cite{RBKSC04}. This has a similar
geometric interpretation of inscribing inside the set ${\cal M}_N$
of mixed states an $N^2-1$  dimensional  simplex spanned by $N^2$
pure states $|\phi_j\rangle$,
 the overlap of which is constant,
 $F=|\langle \phi_i|\phi_j\rangle|^2=1/(N+1)$ for any $i \ne j$.
 Therefore, in this case, the side of the simplex
with respect to the Bures distance reads $D_B^{\rm
SIC}=\sqrt{2(1-\sqrt{F})}= \sqrt{2-2/\sqrt{N+1}}$, and for a finite
dimension $N$, this is smaller than $D_B^{\rm max}=\sqrt{2}$.

So in the distinguishability problem we wish to embed into the set
${\cal M}_N$ of mixed states a simplex of the maximal side length
$D_B^{\rm max}$ with dimensionality not larger than $N$, while in
the SIC POVM problem we try to inscribe inside the same set a higher
dimensional simplex of a smaller side length $D_B^{\rm SIC}$.


\section{Distances between unitary orbits}
\label{SCN: unitary orbits}

In this section we shall be concerned with the distances between
orbits generated from quantum states by unitaries. That is, given
two states $\rho_1$ and $\rho_2$ with fixed spectra, we wish to know
how ``far'' or how ``close'' we can make these states by unitary
action. We will find that for the Bures and trace distance, the closest 
and the farthest that can be achieved is given when both states are diagonal 
in the same basis. This has a geometric interpretation in terms of the Weyl
chambers as will be discussed.

This problem can be interesting in many areas of quantum information. Operationally the problem of
finding the best unitary separation of two density matrices may be interesting if we are restricted to
certain spectra or mixedness. For example in coding for noisy channels. If we know that the output of
some channel will imply a certain mixedness (or even specific spectra), we naturally want to choose to
encode on states that are least affected by this. If we are encoding classical information, this would
be those states which remain most distinguishable afterwards. A simple example of such a channel would
be one which probabilistically adds white noise. Freedom of the input state would correspond to unitary
freedom of the outputs states which we wish to optimise over, hence considering the optimum over
unitary orbits of the output mixed states is equal to finding the optimum encoding. We will see that in
such cases, when only the spectra are restricted, the worst and best cases are given by taking them
diagonal in the same basis.

Consider first two classical, $N$--point, normalised probability
distributions, $p=(p_1,\dots,p_N)$ and $q=(q_1,\dots,q_N)$ such that
$p_i, q_i\ge 0$ and $\sum_i p_i=\sum_i q_i=1$. As earlier, let
$p^{\downarrow}_i$ denote the vector ordered decreasingly, $p^{\downarrow}_i \ge p^{\downarrow}_{i+1}$, 
while let $p^{\uparrow}_i$ represent components of the probability vector in the increasing order: 
$p^{\uparrow}_i \le p^{\uparrow}_{i+1}$.

Any quantum state $\rho_1$ generates an orbit of unitarily equivalent 
states, $U\rho_1 U^{\dagger}$. Two states $\rho_1$ and $U\rho_1 U^{\dagger}$
are sometimes called {\sl geometrically uniform} and they have been recently 
considered in the context of unambiguous discrimination \cite{RLE03,ESH04,He07}.

We are going to discuss another problem of distinguishing states
from two orbits. Consider two diagonal quantum states, $\rho_1={\rm
diag}(p)$ and $\rho_2={\rm diag}(q)$, from which we two orbits of
unitarily equivalent states. We shall analyze the minimal and
maximal distance $D_x$ between the orbits,
\begin{eqnarray}
M(\rho_1,\rho_2):=\max_{U,V} D_x (U\rho_1 U^{\dagger}, V\rho_2 V^{\dagger})=
\max_{W} D_x (\rho_1, W\rho_2 W^{\dagger}), \\
m(\rho_1,\rho_2):=\min_{U,V} D_x (U\rho_1 U^{\dagger}, V\rho_2 V^{\dagger})=
\min_{W} D_x (\rho_1, W\rho_2 W^{\dagger}),
\label{dorbit0}
\end{eqnarray}
since performing maximization over two unitary matrices $U$ and $V$
is equivalent to find a single unitary matrix $W=U^{\dagger} V$.
Here $D_x$ stands for one of the monotone distances $D_{\rm B}$ or
$D_{\rm tr}$. A similar statement for the non monotone
Hilbert-Schmidt distance (\ref{HS1}) was already proved in
\cite{ZB02}.

We conjecture that extrema for these distances are obtained for
diagonal matrices. Then the extremization has to be performed only
over the group $P$ of permutation matrices, which change the order
of the spectra,
\begin{eqnarray}
M(\rho_1,\rho_2)=\max_{P} D_x (p,q)=
  D_x (p^{\downarrow},q^{\uparrow})= D_x (p^{\uparrow},q^{\downarrow}), \\
m(\rho_1,\rho_2)=\min_{P} D_x (p,q)=
  D_x (p^{\downarrow},q^{\downarrow})= D_x (p^{\uparrow},q^{\uparrow}).
\label{dorbit1}
\end{eqnarray}
The minimum is then achieved for the same order of components in both vectors, 
while the maximum occurs for opposite ordering so using the above formula one 
can evaluated analytically the extremal distances for both distances in consideration.

Let us first show that this conjecture holds for the Bures distance.

\begin{theorem}
The maximum and minimum Bures distance between the
unitary orbits of two states are given by diagonal states with
\begin{equation}
M(\rho_1,\rho_2)=\max_{P} D_B (p,q)=
  D_B (p^{\downarrow},q^{\uparrow})= D_B (p^{\uparrow},q^{\downarrow}),
\label{dorbit2a}
\end{equation}
and
\begin{equation}
m(\rho_1,\rho_2)=\min_{P} D_B (p,q)=
  D_B  (p^{\downarrow},q^{\downarrow})= D_B (p^{\uparrow},q^{\uparrow}).
\label{dorbit2}
\end{equation}
\end{theorem}

{\bf Proof. } a) We start by providing an upper bound for the Bures distance (\ref{Bures1}): 

Let us start with the inequality
\begin{eqnarray}
\label{eqn: doub stoch bounds}
\sqrt{p^\uparrow}.\sqrt{q^\uparrow} \ \geq  \
{\rm Tr}\sqrt{\rho_1}\sqrt{\rho_2}    \    \geq \
 \sqrt{p^\uparrow}.\sqrt{q^\downarrow}
\end{eqnarray}
which is a particular case of  (\ref{tracest}) from  lemma
\ref{lem:bound-trace-prod-pow} proved in appendix \ref{app:A}.
Since ${\rm Tr}|\sqrt{\rho_1}\sqrt{\rho_2}| \geq {\rm
Tr}\sqrt{\rho_1}\sqrt{\rho_2}$, we immediately infer
that the root fidelity
is bounded from below by the Bhattacharayya coefficient
between the spectra put in an opposite order,
\begin{equation}
 \sqrt{ F(\rho_1, \rho_2)}=
{\rm Tr}|\sqrt{\rho_1}\sqrt{\rho_2}|
\ge {\rm Tr} \sqrt{\rho_1} \sqrt{\rho_2}
 \ge \sqrt{p^{\uparrow}} \cdot  \sqrt{q^{\downarrow}}
 =   B( p^{\uparrow} \cdot  q^{\downarrow}  )   \ .
\label{Bhat3}
\end{equation}
This implies an upper bound for the Bures distance
which is clearly achievable,
$ M(\rho_1,\rho_2)  = D_B (p^{\uparrow}, q^{\downarrow}).$
 \halmos


In this way we obtain a general upper bound  (\ref{dorbit2a})
for the Bures distance between any two density operators
with spectra $p$ and $q$,
\begin{equation}
D_B(\rho_1,  \rho_2 )  \leq
D_B (p^{\uparrow}, q^{\downarrow})
=  \bigr[ 2(1-\sqrt{p^\uparrow}.\sqrt{q^\downarrow} ) \bigl] ^{1/2} \  .
\label{Buresup}
\end{equation}

b) Next we provide a lower bound for the Bures distance
(\ref{Bures1}):


To prove the case for minimisation our task is to show
\begin{eqnarray}
\label{lower1}
\sqrt{p^\uparrow}.\sqrt{q^\uparrow} \geq {\rm Tr}
|\sqrt{\rho_1}\sqrt{\rho_2}|,
\end{eqnarray}
or equivalently, to get an upper bound for the root fidelity
${\sqrt {F(\rho_1, \rho_2 )}}$.

First we note that for any operator $A$ we have \cite{kf,FG99,HornJohnson2}
\begin{equation}
    \max_U |\tr UA| = \tr\sqrt{AA^\dagger}\equiv\tr|A|\equiv||A||_1,
    \label{eqn:trace-unit-max}
\end{equation}
where the maximum is taken over all unitaries $U$. We will also use
the inequality of von Neumann inequality \cite{VN37}, which concerns 
absolute value of the trace of a product of two matrices and their singular values.


\begin{lemma}[von Neumann inequality]
Let $\sigma_1(A),\ldots,\sigma_n(A)$ and  $\sigma_1(B),\ldots,\sigma_n(B)$
denote singular values of the matrices $A$ and $B$ arranged in
nonincreasing order.  For any matrices $A$ and $B$ the following inequality holds
\begin{equation}
    |\tr AB| \leq \sum_{i=1}^{n}\sigma_i(A)\sigma_{i}(B) \ .
    \label{eqn:vn-tr}
\end{equation}
\end{lemma}
For a recent exposition see \cite{mirsky73trace} and \citep{bourin}.


Without loosing the generality  we can assume that $\rho_1$ is diagonal,
 $\rho_1=\mathrm{diag}(p)$ and
$\rho_2=V \mathrm{diag}(q) V^\dagger$. Then
\begin{equation}
\max_V \sqrt{F}(\rho_1,\rho_2)=\max_V \tr|\sqrt{\rho_1}\sqrt{\rho_2}| =
\max_V \tr |\sqrt{p}V\sqrt{q}V^\dagger|.
\end{equation}
Using (\ref{eqn:trace-unit-max}) and the cyclic property of trace we get
\begin{eqnarray}
\max_V \sqrt{F}(\rho_1,\rho_2) &=& \max_{V,U} |\tr
U\sqrt{p}V\sqrt{q}V^\dagger| = \max_{V,U} |\tr  \sqrt{p} V
\sqrt{q}V^{\dagger}U| \\
&=& \max_{V,W} |\tr  \sqrt{p} V \sqrt{q}W|
\end{eqnarray}
where $W=V^{\dagger}U$ is unitary.
Since the vectors $\sqrt{p}$ and $\sqrt{q}$ contain singular
values of matrices $\sqrt{p}V$ and $\sqrt{q}W$, respectively,
it follows from (\ref{eqn:vn-tr})  that
\begin{equation}
    |\tr \sqrt{p}V\sqrt{q} W| \leq  \sum_{i=1}^{n} \sigma_i^\uparrow( \sqrt{p}V)
    \sigma_i^\uparrow( \sqrt{q} W) \ .
    \label{eqn:vn-aplic1}
\end{equation}
Thus we get the bound for the maximal root fidelity at the unitary orbit,
\begin{eqnarray}
\max_V \sqrt{F}(\rho_1,\rho_2)
&\leq& \sum_{i=1}^{n} \sigma_i^\uparrow(\sqrt{p} V)\sigma_i^\uparrow(\sqrt{q} W) \\
&=& \sqrt{p^\uparrow}\cdot \sqrt{q^\uparrow} \ .
\end{eqnarray}
This result implies the desired upper bound for the root fidelity,
\begin{equation}
    \sqrt{F}(\rho_1,\rho_2)
    \leq \sqrt{p^\uparrow}\cdot \sqrt{q^\uparrow} \ ,
\label{fidup}
\end{equation}
which finishes the proof of the lower bound (\ref{dorbit2}).
Squaring the relations (\ref{Bhat3}) and (\ref{fidup}) we establish the 
inequalities (\ref{fidboth}) and (\ref{burboth}).\halmos


Now we are going to formulate and prove an analogous conjecture for the trace distance.

\begin{theorem}
The maximum and minimum trace distance between the
unitary orbits of two states are given by diagonal states with
\begin{equation}
M(\rho_1,\rho_2)=\max_{P} D_{\rm tr} (p,q)=
D_{\rm tr} (p^{\downarrow},q^{\uparrow})= D_{\rm tr} (p^{\uparrow},q^{\downarrow}),
\end{equation}
and
\begin{equation}
m(\rho_1,\rho_2)=\min_{P} D_{\rm tr} (p,q)=
D_{\rm tr} (p^{\downarrow},q^{\downarrow})= D_{\rm tr} (p^{\uparrow},q^{\uparrow}).
\end{equation}
\end{theorem}

{\bf Proof.} 
The above theorem can be expressed in term of singular values as 
\begin{equation}
  \sum_{i=1}^{n} |\sigma_i(\rho_1)-\sigma_i(\rho_2)|\leq
  \sum_{i=1}^{n}\sigma_{i}(\rho_1-\rho_2) \le
 \sum_{i=1}^{n} |\sigma_i(\rho_1)-\sigma_{n+1-i}(\rho_2)|.
\label{boun2}
\end{equation}
Here $\sigma_i(\rho_1)$ and $\sigma_i(\rho_2)$  denote decreasingly ordered 
singular values of both operators.

The lower bound follows from the special case ($k=n$) of the
following lemma from~\cite{HornJohnson2}.

\begin{lemma}
  Let $A,B\in M_{n}$, and suppose $A, B, A-B$ have decreasingly ordered singular
  values $\sigma_1(A)\geq\ldots\geq\sigma_n(A), \sigma_1(B)\geq\ldots\geq\sigma_n(B), 
  \sigma_1(A-B)\geq\ldots\geq\sigma_n(A-B)$. Define $s_i(A,B)\equiv|\sigma_i(A)-\sigma_i(B)|$
  and let $s_{[1]}(A,B)\geq\ldots\geq s_{[n]}(A,B)$ denote a decreasingly
  ordered rearrangement of the values $s_i(A,B)$. Then
  \begin{equation}
	\sum_{i=1}^k s_{[i]}(A,B) \leq \sum_{i=1}^k\sigma_i(A-B)\ \mathrm{for}\
	k=1,2,\ldots,n.
  \end{equation}
\end{lemma}

The upper bound in (\ref{boun2}) follows from lemma
\ref{lem:upper-trace} in appendix \ref{app:B} if $A$ and $B$ are positive
semidefinite. Since any two density matrices, $\rho_1$ and $\rho_2$, 
are  hermitian and positive, their eigenvalues and singular values are equal.
Making use of the definition (\ref{trace1}) we obtain therefore required
bounds for the trace distance
\begin{equation}
  2 D_{tr}(p^{\downarrow},q^{\downarrow}) 
\   \leq  \  
             {\rm Tr} |\rho_1 - \rho_2|  
\   \le  \ 
  2 D_{tr}(p^{\downarrow},q^{\uparrow})
\end{equation}
equivalent to eq. (\ref{tracboth}).
\halmos

\begin{figure} [htbp]
  \begin{center}
	\includegraphics[width=11.0cm,angle=0]{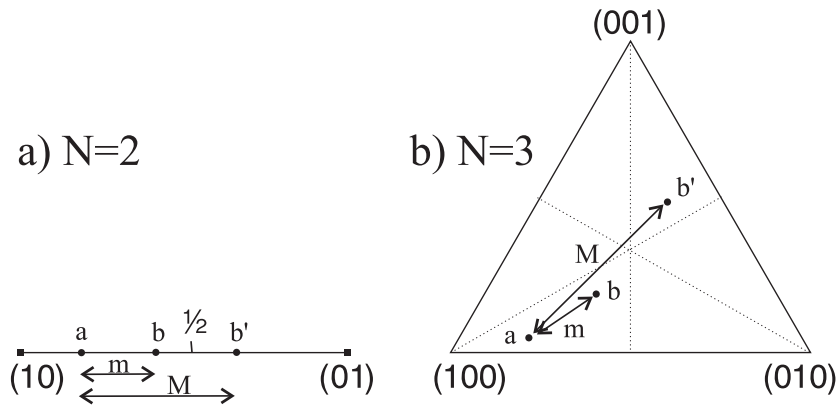}
	\caption{The minimal distance $m$ between the orbits of unitarily similar states stemming from two
	quantum states are equal to the distances between the corresponding spectra $a$ and $b$ belonging to
	the same Weyl chamber shown for a) $N=2$ and b) $N=3$. The maximal distance $M$  is achieved for 
	points $a$ and $b'$ belonging to the opposite Weyl chambers.} \label{fig2}
  \end{center}
\end{figure}

We now consider what this means geometrically, and we will do this
in terms of the so called Weyl chamber. A Weyl chamber is a simplex
of ordered eigenvalues (see, e.g. \cite{BZ06}). Any unitary orbit is
generated from an ordered spectrum of the density matrix, which
corresponds to a point inside a Weyl chamber, i.e. the asymmetric
$1/N!$ part of the simplex of eigenvalues. Thus the minimal distance
between a diagonal state $\rho_1$ and a unitary orbit stemming from
$\rho_2$ is obtained if the orbit intersects the Weyl chamber
distinguished by $\rho_1$. On the other hand the maximum is achieved
also for a diagonal $\rho_2$ with permuted spectrum, which belongs
to another Weyl chamber (see Fig.~\ref{fig2} for $N=2$ and $N=3$).

Let us analyze the simplest case $N=2$, for which the simplex of eigenvalues is equivalent to an
interval $[0,1]$, while the intervals  $[0,1/2)$ and $(1/2,1]$ form two Weyl chambers. A unitary orbit
generated by each point of a Weyl chamber has the structure of the sphere, $S^2$. The above statement
has an intuitive interpretation: the minimal distance between two concentric spheres is equal to the
distance between two of their points belonging to the same radius of the ball. The maximal distance
between these spheres equals to the distance between their points placed at the diameter of the ball on
the other sides of its center. For example consider two states in the Bloch ball. The radius is given
by the entropy, in this case completely defining the spectrum also. So two orbits are given by two
concentric spheres of different radius. Common eigenbases corresponds to a common axis, hence the
closest and furthest states both lie on the same axis, either both on the same side or opposite sides
of the center respectively.

The above property shows that looking for a set of perfectly
distinguishable states in a certain set $S$ of mixed states which is
invariant with respect to the unitary rotations, it is enough to
analyze the subset of diagonal matrices.

\begin{proposition}
Let $R_{\Delta}$ be an arbitrary convex subset of the $(N-1)$ dimensional simplex of the eigenvalues.
Let $R$ denote the set of quantum states obtained from this set by any unitary rotation, 
$R:=\{\rho\in {\cal M}_N:\ \rho=U [{\rm diag }(p)] U^{\dagger}, \ {\rm and}  \ p \in R_{\Delta} \}.$ 
Let $k \le N$ be the number such that $\Delta_{k-1} \in R_{\Delta}$ and there exists no 
$\Delta_{k} \in R_{\Delta}$. Then the maximal  number of perfectly distinguishable states in
$R$ is equal to $k$, so it is equal to the maximal number of diagonal distinguishable states.
\end{proposition}


As before, the symbol  $\Delta_{k}$ represents a regular $k$ dimensional simplex containing $k+1$ 
points separated by the maximal distance $D^{\rm max}$ with respect to the trace (or Bures)
distance. Let us emphasise again that the geometry induced by the Bures metric differs considerably 
with respect to the flat Euclidean geometry induced by the HS metric. For instance, the simplex of
eigenvalues for $N=3$ forms a flat equilateral triangle (of side $\sqrt{2}$) in the HS case, 
while it is equivalent to the octant of a sphere $S^2$ for the Bures distance.

\section{Conclusions} \label{SCN:conclusions}

In this work we have commenced with the analysis of the geometry
 of the problem of quantum distinguishability. We have shown
that the problem of finding the maximal number of perfectly
distinguishable states in a certain set $R$ containing quantum
states is equivalent to finding the dimension of the largest simplex
of a fixed side size which can be embedded inside the set $R$. For
this purpose one cannot use Euclidean simplices defined by the HS
distance, but use simplices with respect to Bures or  trace
distances.

Fidelity between any two quantum states is shown to be bounded
by the classical fidelities between both spectra put in the same 
order (upper bound) or in the opposite order (lower bound). This 
observation implies bounds for the Bures distance between two 
quantum states are achieved for diagonal states. Thus looking for 
distinguishable states in a rotationally invariant subset of the 
set of quantum states it is sufficient to restrict analysis to a smaller set of
classical states, which correspond to diagonal density matrices.

\begin{acknowledgments}
It is a pleasure to thank I. Bengtsson and P. Horodecki for inspiring
discussions and to C.R. Johnson for helpful correspondence.

We acknowledge financial support by the Polish Ministry of Science and Information Technology 
and by the European Research Project SCALA. DM acknowledges support from QICS.

This project was also partially funded by Polish Ministry of Science and Higher Education 
grant number N519 012 31/1957. 
\end{acknowledgments}

\appendix

\section{Bound for the trace of a product of states}\label{app:A}

Let $\rho=\rho^{\dagger}$ and $\sigma=\sigma^{\dagger}$ denote two
Hermitian operators acting on an $N$--dimensional Hilbert space. As
throughout the paper, their spectra will be denoted by $p={\rm
eig}(\rho)$ and $q={\rm eig}(\sigma)$ respectively. Let
$p^{\downarrow}, q^{\downarrow}$ denote the $N$-element vector of
eigenvalues ordered in decreasing order, while the same spectra
ordered increasingly will be written as $p^{\uparrow}$ and
$q^{\uparrow}$. The symbol $(p^{\uparrow})^s$ denotes the vector
consisting of ordered elements of $p^{\uparrow}$, each component
raised to power $s$.


\begin{lemma}\label{lem:bound-trace-prod-pow}
Let $\rho\ge 0$ and $\sigma\ge 0$ and let $s,t$ denote positive 
real numbers. Then
\begin{equation}
  (p^s)^{\uparrow} \cdot (q^t)^{\downarrow}
  \  \le \    {\rm Tr}\, \rho^s \sigma^t  \   \le \
  (p^s)^{\uparrow} \cdot (q^t)^{\uparrow}  \ .
  \label{tracest}
\end{equation}
\end{lemma}


{\bf Proof}. Let $|\mu_i\rangle$ and $|\nu_j\rangle$ denote the eigenvectors of the states $\rho$ and
$\sigma$. We will start by finding a form of ${\rm Tr}\rho^s {\sigma}^t$,
 in terms of overlaps with a doubly stochastic matrix.

\begin{eqnarray}
{\rm Tr} \rho^s \sigma^t &=& {\rm
Tr}\left(\sum_{i,j} p_i^s q_j^t |\mu_i\rangle\langle
\mu_i|\nu_j\rangle\langle\nu_j|\right) \\
&=&\sum_{i,j}p_i^s q_j^t |\langle \mu_i|U|\mu_j\rangle|^2\\
&=&\sum_{i,j} p_i^s q_j^t B_{i,j},
\end{eqnarray}
where $U$ is the unitary relating the two eigenbases
$U|\mu_i\rangle=|\nu_i\rangle$, $\forall i$ and
$B:=\sum_{i,j}|U_{i,j}|^2|\mu_i\rangle\langle\mu_j|$
so that $B_{ij}=|U_{ij}|^2$.
Hence matrix $B$ is by construction unistochastic \cite{ZKSS03}
and thus bistochastic.

It is convenient to introduce at this place two non-normalised
vectors, $|\psi\rangle:=\sum p_i' |\mu_i\rangle$, and
$|\phi\rangle:=\sum q_j' |\mu_j\rangle$, where $p_i'=p_i^s$ and
$q_j'=q_j^t$ are non--negative. Then the trace can be rewritten in
the form
\begin{equation}
{\rm Tr} \rho^s \sigma^t = \langle \psi |B| \phi\rangle.
\label{psiMphi}
\end{equation}
Birkhoff's theorem \cite{HornJohnson} states that any doubly
stochastic matrix can be written as a finite convex combination of
permutation matrices $O_i$, hence we write $B=\sum_i r_i O_i$,
$\sum_i r_i =1$.
Thus the extremum of a linear function of the bistochastic matrix $B$
will be realized at one of its extremal points. There are exactly $N!$
of them, and among all possible permutations $O_i$
the maximum is obtained if
the orders of elements of both vectors
are the same, while the minimum
is achieved if  both spectra are in opposite order,
\begin{eqnarray}
\langle \psi |B|\phi\rangle = \sum_i r_i \langle \psi
|O_i|\phi\rangle & \geq & \langle \psi |O_{min}| \phi\rangle =
(p^{\uparrow})^s \cdot (q^{\downarrow})^t \\
& \leq & \langle \psi |O_{max}|\phi\rangle =
(p^{\uparrow})^s \cdot (q^{\uparrow})^t \ .
\label{psiMphi2}
\end{eqnarray}

Since all components of the vector $p$ (and $q$) are non--negative
raising each element to a positive exponent $s$ (or $t$)
will not change the order of a vector, $(p^{\uparrow})^s =(p^s)^{\uparrow}$.
Putting it all together we arrive at (\ref{tracest})
and complete the proof. \halmos


For concreteness let us write down explicitly some special cases. In the simplest case $s=t=1$ one
obtains
\begin{equation}
\label{eqn:bounds11}
p^{\uparrow} \cdot q^{\downarrow}
\ \leq  \ {\rm Tr} \rho \sigma  \  \leq \
p^{\uparrow} \cdot q^{\uparrow} \ ,
\end{equation}
while setting $s=t=1/2$ one becomes inequality
(\ref{eqn: doub stoch bounds})
used in the proof of inequality (\ref{Buresup}).

An analogue of lemma 2 may be obtained
in the case one of the two operators is not positive.

\begin{lemma}
Consider a positive number $s>0$
  a state $\rho\ge 0$ and an Hermitian operator $\sigma=\sigma^{\dagger}$
not necessarily positive. Then
 \begin{equation}
(p^s)^{\uparrow} \cdot q^{\downarrow}
\  \le \    {\rm Tr}\rho^s \sigma  \   \le \
(p^s)^{\uparrow} \cdot q^{\uparrow}  \ .
\label{tracest2}
\end{equation}
\end{lemma}

Proof of this lemma is similar to the proof of lemma 2.
In this case the vector $q$ of eigenvalues of operator $\sigma$
contains in general also negative entries, so the
vector $|\phi\rangle:=\sum_j q_j |\mu_j\rangle$, is given by a pseudomixture
with some weights negative. Constructing unitary bases $U$ and bistochastic 
matrix $M$ one may write the analyzed trace in the form (\ref{psiMphi})
and make use of the Birkhoff theorem. Since operator $\rho$ with spectrum $p$ 
is positive, raising its components to a positive power will not change the 
order, $(p^{\uparrow})^s =(p^s)^{\uparrow}$. Therefore we may perform the last 
step analogous to (\ref{psiMphi2}) obtaining the desired result. \halmos


\section{Bound for the trace of a difference of two states}\label{app:B}

In this appendix we prove the following lemma.
\begin{lemma}\label{lem:upper-trace}
Let $A$ and $B$ denote hermitian matrices of size $n$.
Let us order decreasingly their eigenvalues, $\lambda_1(A)\geq\ldots\geq\lambda_n(A)$
and $\lambda_1(B) \geq \ldots \geq \lambda_n(B)$. Then the following upper
bound for the trace of the absolute value of the difference holds 
\begin{equation}
 {\rm Tr}|A-B| \ = \   \sum_{i=1}^{n}\sigma_{i}(A-B) 
\  \le  \ 
  \sum_{i=1}^{n} |\lambda_i(A)-\lambda_{n+1-i}(B)|.
\label{newlemma}
\end{equation}
\end{lemma}

{\bf Proof.} 
Let us express both operators
in their eigen representation, 
$A=\sum_i^n p_i |\mu_i\rangle \langle \mu_i|$ and 
$B=\sum_i^n q_i |\nu_i\rangle \langle \nu_i|$,
where for convenience we have introduced the notation
$p_i=\lambda_i(A)$ and $q_i=\lambda_i(B)$. 
Making use of  Eq.~(\ref{eqn:trace-unit-max}) 
and basic properties of the trace we get
\begin{eqnarray}
\tr|A-B| & = & \max_U|\tr AU - \tr BU |\\
& = & \max_U|\sum_{i=1}^n p_i \bra{\mu_i}U\ket{\mu_i} - q_i
\bra{\nu_i}U\ket{\nu_i}|.
\end{eqnarray}
Since $|\bra{\mu_i}U\ket{\mu_i}| \leq 1$, $|\bra{\nu_i}U\ket{\nu_i}| \leq 1$ and 
$\tr U = \sum_{i=1}^n \bra{\mu_i}U\ket{\mu_i} = \sum_{i=1}^n  \bra{\nu_i}U\ket{\nu_i}$ we have
\begin{eqnarray}
\tr|A-B|  &\leq& \max 
\left\{
	|\sum_{i=1}^n \xi_i  p_i - \zeta_i q_i | :  
	|\xi_i | \leq 1 , |\zeta_i | \leq 1 
	\text{ for } i = 1, \dots , n \; ,\sum_{i=1}^n \xi_i = \sum_{i=1}^n \zeta_i
\right\}.
\end{eqnarray}
For fixed values of $\xi_i$ and
$\zeta_i$ we denote $s = \sum_{i=1}^n \xi_i  p_i - \zeta_i q_i$. Let $s = c e^{i
\varphi}$, we have
$$
c = |s| = |\frac{s}{e^{i \varphi}}| = |\sum_{i=1}^n \frac{\xi_i}{e^{i
\varphi}}  p_i - \frac{\zeta_i}{e^{i \varphi}} q_i |.
$$
Because  
$|\frac{\xi_i}{e^{i \varphi}}| \leq 1$ and 
$|\frac{\zeta_i}{e^{i \varphi}}| \leq 1$
we can without loss of generality assume that $s \in \bbR$. 
Note now that under this assumption we have
\begin{eqnarray}
&&\max 
\left\{
	|\sum_{i=1}^n \xi_i  p_i - \zeta_i q_i| :  
	|\xi_i | \leq 1 , |\zeta_i | \leq 1 
	\text{ for } i = 1, \dots , n \; ,\sum_{i=1}^n \xi_i = \sum_{i=1}^n \zeta_i,
\right\} \\
&=&\max 
\left\{
	|\sum_{i=1}^n Re(\xi_i)  p_i - Re(\zeta_i) q_i | :  
	|\xi_i | \leq 1 , |\zeta_i | \leq 1 
	\text{ for } i = 1, \dots , n \; ,\sum_{i=1}^n \xi_i = \sum_{i=1}^n \zeta_i,
\right\} \\
&=&\max 
\left\{
	|\sum_{i=1}^n \xi_i  p_i - \zeta_i q_i | :  
	-1 \leq \xi_i \leq 1 ,-1 \leq \zeta_i  \leq 1 
	\text{ for } i = 1, \dots , n \; ,\sum_{i=1}^n \xi_i = \sum_{i=1}^n \zeta_i,
\right\} 
\end{eqnarray}

The term $\sum_{i=1}^n \xi_i  p_i - \zeta_i q_i$ is a linear function of $2 n$ variables $\xi_1, \dots \xi_n , \zeta_1, \dots , \zeta_n$, 
so it reaches its extreme value at the edges of the polygon defined by
\begin{equation}
\left \{
	-1 \leq \xi_i \leq 1 ,-1 \leq \zeta_i  \leq 1 
	\text{ for } i = 1, \dots , n \; ,\sum_{i=1}^n \xi_i = \sum_{i=1}^n \zeta_i
\right \}.
\end{equation}
Thus we can focus on the edges of the polygon

\begin{equation}
\left \{
	\xi_i \in \{-1,1\}, \zeta_i \in \{-1,1\}  
	\text{ for } i = 1, \dots , n \; ,\sum_{i=1}^n \xi_i = \sum_{i=1}^n \zeta_i
\right \}.
\end{equation}

Note that we obtain the maximum if in the sum $\sum_{i=1}^n \xi_i  p_i + (-
\zeta_i) q_i$ the $n$ maximum values of $\{p_1, \dots , p_n, q_1 , \dots ,
q_n\}$ will be equipped with $+1$ coefficient and $n$ minimum values with
$-1$.
Because $p_1 \geq p_2 \geq \dots \geq p_n$ and $q_1 \geq q_2 \geq \dots \geq q_n$, we can thus write the $n$ maximum values  as 
\begin{equation}
\max\{p_1,q_n\},\max\{p_2,q_{n-1}\}, \dots , \max\{p_n,q_1\},
\end{equation}
and the $n$ minimum values as
\begin{equation}
\min\{p_1,q_n\},\min\{p_2,q_{n-1}\}, \dots , \min\{p_n,q_1\}.
\end{equation}
So the maximum value of 
\begin{equation}
\max 
\left\{
	|\sum_{i=1}^n \xi_i  p_i - \zeta_i q_i | :  
	\xi_i \in \{-1,1\}, \zeta_i \in \{-1,1\}  
	\text{ for } i = 1, \dots , n \; ,\sum_{i=1}^n \xi_i = \sum_{i=1}^n \zeta_i
\right\}
\end{equation}
is equal
\begin{equation}
|\sum_{i=1}^n \max\{p_i,q_{n-i+1}\} - \min\{p_i,q_{n-i+1}\}| = \sum_{i=1}^n |p_i - q_{n-i+1}|.
\end{equation}

This gives us required upper bound  (\ref{newlemma}).
\halmos


\end{document}